# STIMULATED LOW-FREQUENCY RAMAN SCATTERING IN ALBUMIN

*M.A. Shevchenko[1], L.L. Chaikov[1], M.N. Kirichenko[1], A.D. Kudryavtseva[1]\*, T.V. Mironova[1], V.I. Savichev[1,2], V.V. Sokovishin[2], N.V. Tcherniega[1], K.I. Zemskov[1]*

[1]P.N. Lebedev Physical Institute of the Russian Academy of Sciences, Moscow, Russia, 119991

[2] Bauman Moskow State Technical University, Moscow, Russia, 105005

**Abstract**

Stimulated low-frequency Raman scattering (SLFRS) can give essential information about the elastic properties of different nanoparticles systems, in particular, biological nanostructures. In the present study, low-frequency vibrational modes in human and bovine serum albumin (HSA and BSA) were for the first time investigated using SLFRS method. 20 ns ruby laser pulses have been used for excitation. SLFRS frequency shifts, corresponding to acoustic eigenfrequencies of the sample, were registered by Fabri-Perot interferometers. For HSA obtained set of eigenfrequencies was 6 GHz (0.2 cm$^{-1}$), 10 GHz (0.33 cm$^{-1}$) and 15.6 GHz (0.52 cm$^{-1}$), for BSA – 8.7 GHz (0.29 cm$^{-1}$) and 16.5 GHz (0.55 cm$^{-1}$). Conversion efficiency and threshold were also measured. SLFRS can be applied for biological objects identification and impact on them.

.

## 1. Introduction

Low-frequency Raman scattering (LFRS) is a powerful and convenient tool for the investigation of various objects with size from several nanometers to micron [1-3]. In the process of inelastic light scattering the frequency shifts corresponding to eigenfrequencies of acoustic vibrations are registered. Since eigenfrequency is determined by the morphology of the objects under study, such important characteristics as size and sound velocity can be determined from the LFRS spectra. Many biological objects of interest are in this size range with eigenfrequencies lying in the gigahertz range (for example viruses and proteins).

Viruses consist of protein shell (capsid) containing genetic material inside (RNA or DNA). Capsids of the most viruses have a spherical or cylindrical form with dimensions from 10nm to 100nm. Viral capsids are flexible and change size or shape in response to vibrations or changes in external conditions.[4] There are currently many different possible applications of viruses in such areas as optoelectronics and nanotechnology[5,6]. The structure of viruses has been well studied by x-ray crystallography [7] in contrast to their dynamic properties. So the study of these properties is an essential task since morphology of viruses significantly depends on their environmental conditions. There are a number of works in which acoustic vibrations of some viruses have been investigated using LFRS methods [8-10]. If the acoustic mode satisfies the selection rules, it is Raman-active and appears in the scattering spectrum as a shifted component [11]. There are two approaches to calculate the discrete spectrum of the acoustic eigenfrequencies of spherical viruses - an elastic sphere model [12] and a liquid drop model [13].

In elastic sphere model the frequency of the normal mode ω=ξV/r (1), where V is sound velocity, r is the radius of the virus particle, ξ is a dimensionless parameter depending on the relationship between longitudinal and transverse sound velocities. In the case of a liquid drop model the frequency of the lowest vibrational mode

$$\omega = \sqrt{\frac{\gamma}{\rho r^3}}$$

, where γ is surface tension, ρ is material density and r is a radius of the virus particle.

In [13] Ford examined both of these models for a spherical virus particle and concluded that the elastic sphere model better describes a virus particle. In any case, the question of choosing a model remains open.

In [14] Babincová with co-workers suggested a way of human immunodeficiency virus (HIV) destruction in the process of ultrasound resonant absorption in GHz range.

Balandin and Fonoberov [15] for the first time calculated the lowest vibrational modes of tubular viruses (M13 bacteriophage and tobacco mosaic virus TMV) in water and air. First experimental results for the low vibrational modes of M13 phages have been obtained in [10]. The observed Raman mode has been shown to belong to one of the Raman-active axial torsion modes of the M13 phage protein coat. It was also theoretically and experimentally shown that only the axial modes are observable in the Raman experiment whereas the radial modes are strongly damped due to the radiation of the acoustic energy into the environment.

D. Murray and L. Saviot [16, 17] also showed the influence of viscosity of water on the oscillation frequency of viruses in water. M. Talati and P. Jha [8] investigated the frequency shift and Raman peak broadening due to damping when the virus is immersed in water and glycerol. The study has been also made at different temperatures. The low-frequency Raman peaks for a water system were more broadened than those for a glycerol one. The authors associate it with the formation of stronger hydrogen bonds at the viral protein surface in case of polar water medium. Estimated damping time for water was two times faster than for glycerol.

Thus, as it was shown in the works mentioned above, for nanoscale viruses both the frequency and damping of the vibrational modes are significantly affected by the properties of a surrounding medium.

So considering that spontaneous scattering mode is characterized by low scattered radiation intensity, using of its stimulated analog (SLFRS) seems more promising for vibrational dynamic study. Stimulated scattering, excited with short and powerful laser pulses, is characterized by high conversion efficiency. It is also possible to apply it as a source of biharmonic pumping with the difference frequency tuning range from a few gigahertzes to terahertz. As a result, we have the vibrational modes resonance excitation that can be used as an effective method of influencing the virus system (like on any biological system). SLFRS was realized for a large number of various nanoparticles (metallic, dielectric, and semiconductor), nanostructured thin films and highly ordered samples such as opal matrices and nanocomposites on their base [18-21]. SLFRS was observed also in different viruses (Cauliflower mosaic virus, Tobacco mosaic virus and two types of Potato viruses) [22, 23, 24]. Cauliflower mosaic virus (CaMV) has a spherical form with a diameter of 35 nm. Tobacco mosaic virus (TMV) and Potato viruses (PVA and PVX) are cylindrical with lengths of 300 nm, 730 nm, and 715 nm respectively, and diameters of 18 nm, 15 nm, and 13.5 nm respectively. The measurements were made in Tris-HCl pH7.5 buffer and in water. SLFRS frequency shifts, conversion efficiency and threshold in viruses have been measured for the first time. Experimentally obtained frequency shift corresponding to an acoustic eigenfrequency of TMV was 60 GHz, which coincides with radial breathing mode frequency calculated in [15]. The maximum conversion efficiency of pumping light into the SLFRS in viruses was 10 % and the SLFRS threshold was about 0.1 GWcm$^{-2}$.

In this work we for the first time registered SLFRS in human and bovine serum albumins (HSA and BSA), which play an important role in biological processes, being the main serum transport protein.

## 2. Experiment

Albumin is a polypeptide chain forming a globule with a size of approximately 8 nm. HSA and BSA are homologous and differ only in some amino acid residues. They play an essential role in blood plasma, definable by the wide variety of functions of these proteins. Both albumins are nowadays well studied and it's well known that they may form associates with a size of hundreds of nanometers. Because of widespread use of albumins, the task arises of measuring the parameters of monomers and formed aggregates. In our experiment, we used albumin water solutions with pH value equal to 7. The mass concentration of serum albumin in the solution was 10 percent. We used dynamic light scattering (DLS) for size distribution measurement. DLS was realized by a Zetasizer Nano ZS analyzer (Malvern). Before DLS measurements were processed in an ultrasonic bath. Figure 1 shows a radius distribution of HAS aggregates in 10% water solution by dynamic light scattering. As one can see from the figure, besides monomer with size 8 nm, aggregates with radius 50 nm were present in solution. The aggregates of approximately the same radius were present in the BSA solution. The radius distribution of BSA is shown in Figure 2.

Experimental setup for SLFRS measurements in albumin is shown in Figure 3. Ruby laser ($\lambda$ = 694.3 nm, $\tau$ = 20 ns, $E_{max}$ = 0.3 J, $\Delta\nu$ = 0.015 cm$^{-1}$, divergence 3.5·10$^{-4}$ rad) radiation was focused at the center of the 1 cm quartz cell (5) with sample by the lens (4) with focal length 5 cm. Spectral shifts were measured with the help of Fabri-Perot interferometers (6) with different ranges of dispersion from 0.3 cm$^{-1}$ to 8.3 cm$^{-1}$ (9-250 GHz). Simultaneously with spectral measurements, the efficiency of conversion and the SLFRS threshold were measured. SLFRS was excited in 10% solution of HSA when the laser intensity reached a threshold at room temperature. The maximum conversion efficiency of pumping light into the scattered light for HSA was 55 percent. The SLFRS threshold for scattering propagating in backward and forward directions was about 0.1 GWcm$^{-2}$. For the used intensity levels of the exciting radiation, the scattering excitation threshold in BSA at room temperature was not reached. For BSA the threshold was reached only at liquid nitrogen temperature (77K). The experimental spectra and frequency shifts for both samples are shown in Figure 4 and Figure 5.

For HSA obtained set of frequency shifts was 6 GHz (0.2 cm$^{-1}$), 10 GHz (0.33 cm$^{-1}$) and 15.6 GHz (0.52 cm$^{-1}$), for BSA- 8.7 GHz (0.29 cm$^{-1}$) and 16.5 GHz (0.55 cm$^{-1}$). Frequency values are just slightly different for HSA and BSA, which is connected with their similar structure. Due to the rather complex morphology of these proteins, there is no physical model that could be used for calculating their oscillatory dynamics. Each SLFRS spectral line corresponds to the different types of the acoustic eigenvibrations of aggregates constituting this system. Elastic sphere model can be used to estimate the relationship of the value of the frequency shift with the elastic characteristics of the submicron aggregates making up the system under study. Taking into account the Lamb's theory dealing with the free vibrations of a homogeneous spherical elastic body under stress-free boundary condition [12], and assuming the speed of sound in a protein as 1550 ms$^{-1}$ [25] one can conclude that these frequencies are related with the aggregates with an average size of hundred nanometers.

## 3. Conclusions

Analysis of the low-frequency spectrum of the inelastically scattered light in different biological structures like in any nanoparticles system can give essential information about their mechanical properties and can be used for their identification. All SLFRS spectral components correspond to the different vibrational modes of the object. Using SLFRS for the experimental definition of eigenfrequency values one can determine the size of the nanoparticles or sound velocity.

Besides identification SLFR-S in different biological objects can be used as the source of biharmonic pumping for investigating the systems with eigenfrequencies in gigahertz range and also for a powerful impact on such systems. It is well known that albumin's conformation, as well as viral capsids, is highly dependent on various external conditions, like pH, temperature, and others. So, changing these external parameters, we can vary the frequency shift between laser excitation and the Stokes component. If one sends this radiation at the object under study, in the case of exact coincidence of the frequency shift with the object eigenfrequency, the resonant impact on the system can be realized. The use of electromagnetic radiation of the visible or near-infrared range has an advantage over other ways of possible resonant impacts on biological systems - resonant microwave absorption and acoustic radiation in giga- or terahertz range. It's connected with strong absorption of electromagnetic radiation in the microwave range in water and a small length of the propagation of acoustic waves of gigahertz range.

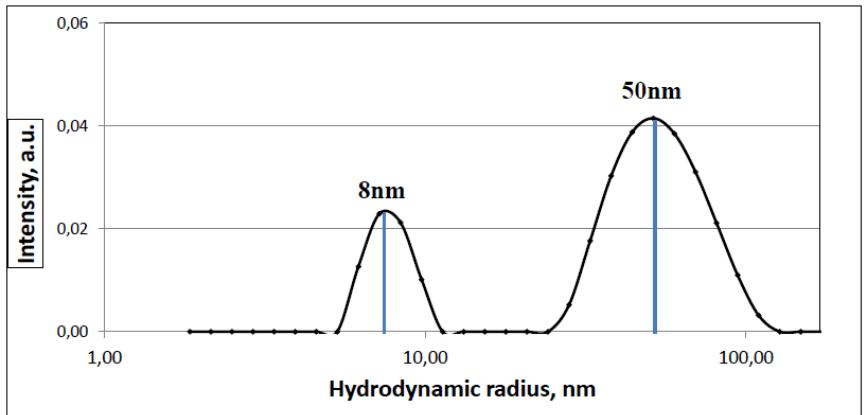
Figure 1: Distribution of hydrodynamic radius of HSA aggregates in 10% water solution.

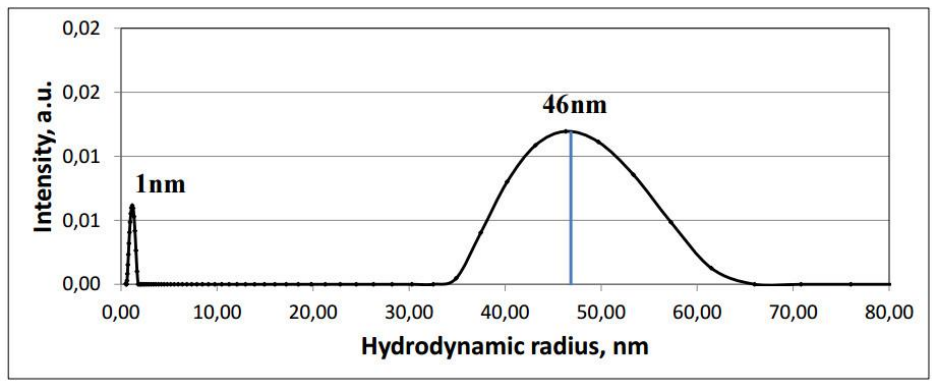

Figure 2: Distribution of hydrodynamic radius of BSA aggregates in 10% water solution.

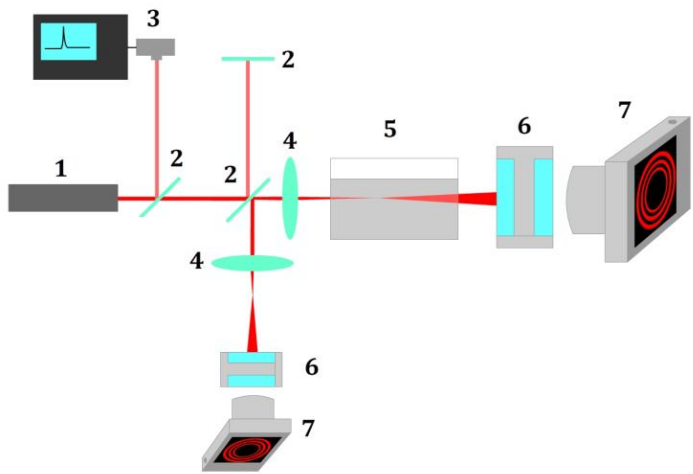

Figure 3: Experimental setup: 1 – ruby laser; 2– glass plates; 5 – quartz cell with the sample; 4– lenses; 6 – Fabri-Perot interferometers; 7 - photo cameras, registering SLFRS spectra; 3 – system for laser pulse characteristics measurements.

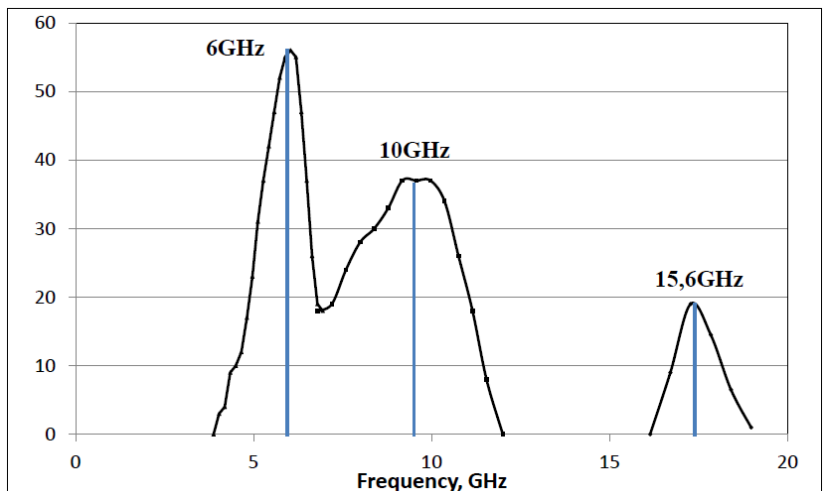

Figure 4: Stimulated low-frequency Raman spectrum of HAS at room temperature.

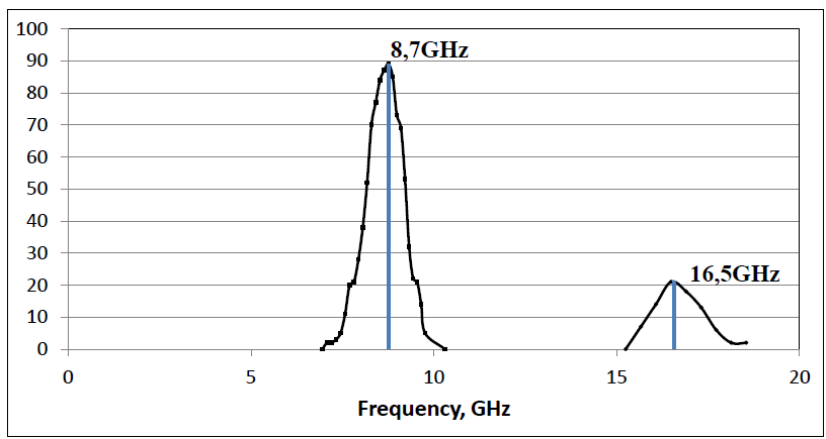

Figure 5: Stimulated low-frequency Raman spectrum of BSA at liquid nitrogen temperature (77K).